\documentclass{article}
\usepackage{tex-style/spconf}
\usepackage{cite}
\usepackage{amsmath,amssymb,amsfonts}
\usepackage{algorithmic}
\usepackage{graphicx}
\usepackage{textcomp}
\usepackage{xcolor}
\usepackage{amsmath}
\usepackage[]{algorithm2e}
\usepackage{diagbox}
\usepackage{todonotes}
\usepackage{blindtext}
\usepackage{caption}
\usepackage{subcaption}
\usepackage{hyperref}

\title{SVT-AV1 Encoding Bitrate Estimation Using Motion Search Information}
\name{
    \parbox{\linewidth}{\centering
    Lena Eichermüller$^{\star}$ 
    \qquad Gaurang Chaudhari$^{\dagger}$
    \qquad Ioannis Katsavounidis$^{\dagger}$  
    \qquad Zhijun Lei$^{\dagger}$ \quad
    \qquad Hassene Tmar$^{\dagger}$
    \qquad Christian Herglotz$^{\star}$
    \qquad André Kaup$^{\star}$ 
    }}

\address{$^{\star}$ Multimedia Communications and Signal Processing \\ 
    Friedrich-Alexander-Universität Erlangen-Nürnberg, Erlangen, Germany \\
    $^{\dagger}$ Meta \\ California, USA}
      
\begin{document}

\maketitle
\begin{abstract}
    Enabling high compression efficiency while keeping encoding energy consumption at a low level, requires prioritization of which videos need more sophisticated encoding techniques.
    However, the effects vary highly based on the content, and information on how good a video can be compressed is required. 
    This can be measured by estimating the encoded bitstream size prior to encoding.
    We identified the errors between estimated motion vectors from Motion Search, an algorithm that predicts temporal changes in videos, correlates well to the encoded bitstream size. Combining Motion Search with Random Forests, the encoding bitrate can be estimated with a Pearson correlation of above $0.96$.
\end{abstract}

\section{Introduction}
Video data dominates the online data traffic and with video streaming becoming ubiquitous in modern entertainment and communication, it is expected to grow even further~\cite{cisco2020cisco}.
However, this demand for online video content comes at a considerable environmental cost, as the storage, transmission, and streaming of videos contributes substantially to global carbon dioxide emissions. A total of $1.3 \%$ of those were caused by video streaming in 2015~\cite{stephens2021carbon}.
Efforts to mitigate these environmental impacts have focused on improving video encoding techniques to reduce the overall data volume required for streaming.
Nevertheless, effective resource management in video streaming systems is crucial for minimizing energy consumption and therefore its associated carbon footprint without diminishing the quality of the viewing experience. 
Prioritizing which videos need more sophisticated encoding techniques is essential in achieving this balance. 
However, determining the optimal encoding strategy for each video poses a challenge on itself, as the effectiveness of encoding techniques varies depending on the content in the video.
One parameter that is affected by the content is the size of the encoded bitstream.
Highly spatially varying content requires more bits for encoding whereas flat areas can be represented by a lower number of bits. 
Also, the movement within a scene influences the required number of bits to encode. 
The storage demand of the video, as well as the transmission of it, is affected by the bistream size.
Hence being able to predict bitstream sizes, or bitrates, of a video for certain sets of parameter configuration prior to encoding, will help for optimizing resource allocation. 
Crucial is an accurate prediction without encoding and also with a significantly lower complexity than the encoding process itself.
This enables the prediction of bitrate ladders for HTTP adapative streaming, which is currently obtained with the help of ultrafast encodings~\cite{de2016complexity,katsenou2021efficient}. 

In the literature, most bitrate models are used during encoding: since not every region within a video frame contains the same amount of detail, the information, or entropy, of the blocks in a video can be estimated during encoding. 
Which blocks require more or less bits to be represented in order to be visually appealing can then decided based on this entropy estimation. 
For example, Sarwer et al. propose a method for bitrate estimation during encoding~\cite{sarwer2007fast}. 
To avoid entropy encoding, the bitrate of 4x4 blocks is estimated during mode decision.

In our work, the goal is to predict the encoded bitrate of a video sequence on a higher level.
The bitrate shall be predicted for an entire video shot and with a significantly lower complexity than the encoding procedure itself for AV1-encoding~\cite{chen2018av1}.
Related work from Haseeb et al. use spatial information (SI) and temporal information (TI) as complexity measures, which has been introduced by~\cite{fenimore1998perceptual}, to model the rate-distortion costs for Scalable Video Coding in real-time~\cite{haseeb2012rate}. 
An improvement to spatial and temporal information is provided by the Video Complexity Analyzer (VCA), and is called \textit{spatial complexity} and \textit{temporal complexity}~\cite{vca}. 
In terms of video coding, VCA shows increased performance for predicting encoding time rather than SI and TI~\cite{amirpour2022light, eichermuller2024encoding} as well as for bitrates~\cite{amirpour2022light} and is used by Amirpour et al. for encoding time and bitrate estimation of H.256-encoding~\cite{amirpour2022light}. 
Coupling it with CNNs and gradient boosting, they obtain a prediction error of $3.47\%$ for the bitrate model, however, only for H.265-encoding.

In this paper, we address bitrate estimation by proposing an approach that combines a Motion Search algorithm with lightweight regression models to estimate the encoded bitrate of videos~\cite{motionsearch2021}. 
We use a Motion Search algorithm that predicts temporal changes in videos, i.e. by estimating motion vectors, and have identified that the errors acquired from those motion vectors correlate well with the encoded bitstream size. 
Adding a low complexity machine learning regressor or creating an analytical model from those features, we develop a bitrate estimator with a high correlation to the encoded bitrate.
First, we analyze how values provided by the Motion Search procedure correlate to the bitrate for AV1-encoding. 
Then, we create a bitrate model via an analytical regression model and Random Forest regression. 
We conclude that using Motion Search instead of VCA increases the Pearson correlation between prediction and actual bitrate from $0.89$ to $0.95$ in the case of AV1-encoding.
We finally construct a model that can be furthermore adjusted to also include temporal and spatial complexity which further increases the correlation coeffcient to $0.96$. 

\section{Video Complexity Estimation via Motion Search}
In order to estimate the temporal changes in a video sequences, we use \textit{Motion Search}~\cite{motionsearch2021}. 
Here, depending on if the current frame is an I-,P- or B-frame, \textit{spatial search}, \textit{motion search}, or \textit{bidirectional search}, respectively, are applied in order to estimate the \textit{number of bits} required to encode the frame, as well as the block-based variance or mean squared error (\textit{MSE}) from calculated motion vectors. 
For this work we ignore B-frames and concentrate on I- and P-frames only.



\subsection{Spatial Search and Motion Search}
Each frame is subdivided in blocks. 
Starting from the upper left, the blocks of the picture are processed in raster-scan error, i.e. row-wise from left to right. 
For blocks in I-frames, called I-blocks, \textit{spatial search} is performed. 
First, the block-based variance 
\begin{equation}
    \mathrm{var}_\mathrm{block} = \sum_{\mathrm{block}} I^2 - \left( \sum_{\mathrm{block}} I \right)^2
    \label{eq:block-var}
\end{equation}
is calculated for blocks of size $16 \times 16$ and blocks of size $8 \times 8$ -- indicated by $\mathrm{block}$ -- over a frame $I$.  
For the $8 \times 8$ block variance $\mathrm{var}_{8 \times 8}$, next to the variance of the current block $\mathrm{var}_{8 \times 8}^{(0)}$, the variances of the neighboring $8 \times 8$ blocks $\mathrm{var}_{8 \times 8}^{(j)}, j \in \left( 1, 2, 3 \right)$ are also taken into account.
We chose the mean-squared spatial block error $\mathrm{MSE}_\mathrm{block}^I$ as the minimum of $16 \times 16$ block variance and $8 \times 8$ block variances, i.e.,
\begin{equation}
    \mathrm{MSE}_\mathrm{block} = \min \left\{ \sum_{i=1}^{4} \mathrm{var}_{8 \times 8}^{\left( i \right)}, \mathrm{var}_{16 \times 16} \right\}.
    \label{eq:mse-block}
\end{equation}
The bitsize $k$ of this block is determined as:
\begin{equation}
    k = \lceil \log_2 \left( \mathrm{MSE}_\mathrm{block}\right)\rceil
    \label{eq:bitsize}
\end{equation}
For P-blocks, we perform \textit{motion search}.
Here, the spatial block error defined as $\mathrm{MSE}_\mathrm{block}^I$ is calculated from \eqref{eq:mse-block} is first.
Motion estimation is performed afterwards and the block-based variances from the motion vector blocks $\mathrm{MSE}_\mathrm{MV, block}$ are calculated similarly to~\eqref{eq:mse-block}:
\begin{equation}
    \mathrm{MSE}_\mathrm{MV, block} = \min \left\{ \sum_{i=1}^{4} \mathrm{var}_{\mathrm{MV}, 8 \times 8}^{\left( i \right)}, \mathrm{var}_{\mathrm{MV}, 16 \times 16} \right\}
\end{equation}
If this value is larger than $\mathrm{MSE}_\mathrm{block}^I$, the block is considered an I-block, else a P-block. 
The resulting motion block error $\mathrm{MSE}_\mathrm{block}$ is again the minimum over all block-variances:
\begin{equation}
\mathrm{MSE}_\mathrm{block} = \min \left\{ \mathrm{MSE}_\mathrm{block}^I, \mathrm{MSE}_\mathrm{MV, block} \right\}
    \label{eq:mse-block-2}
\end{equation}
Analogously to~\eqref{eq:bitsize}, the bitsize of a block is obtained from the spatial block error $\mathrm{MSE}_\mathrm{block}^I$.

\subsection{Content Complexity Descriptors}
We define $\mathrm{MSE}$ as the sum of all block errors $\mathrm{MSE}_\mathrm{block}$ in one frame $I$. 
For the Motion Search error complexity descriptor, we get
\begin{equation}
    \mathrm{MSE}_\mathrm{MS} = \frac{1}{N\cdot M \cdot n_\mathrm{frames}} \cdot \sum^{n_\mathrm{frames}} \mathrm{MSE}(I),
    \label{eq:mse-ms}
\end{equation}
where the accumulated error is normalized by the number of frames $n_\mathrm{frames}$, the height $M$ and the width $N$ of the video.
Similarly, we obtain the estimated bits per pixel:
\begin{equation}
    \mathrm{bpp}_\mathrm{MS} = \frac{1}{N\cdot M \cdot n_\mathrm{frames}} \cdot \sum^{n_\mathrm{frames}} \sum^{n_\mathrm{blocks}}  k(I, \mathrm{block})
    \label{eq:mse-bpp}
\end{equation}   
\begin{figure*}
    \begin{subfigure}[b]{0.245\textwidth}
        \includegraphics[width=\textwidth]{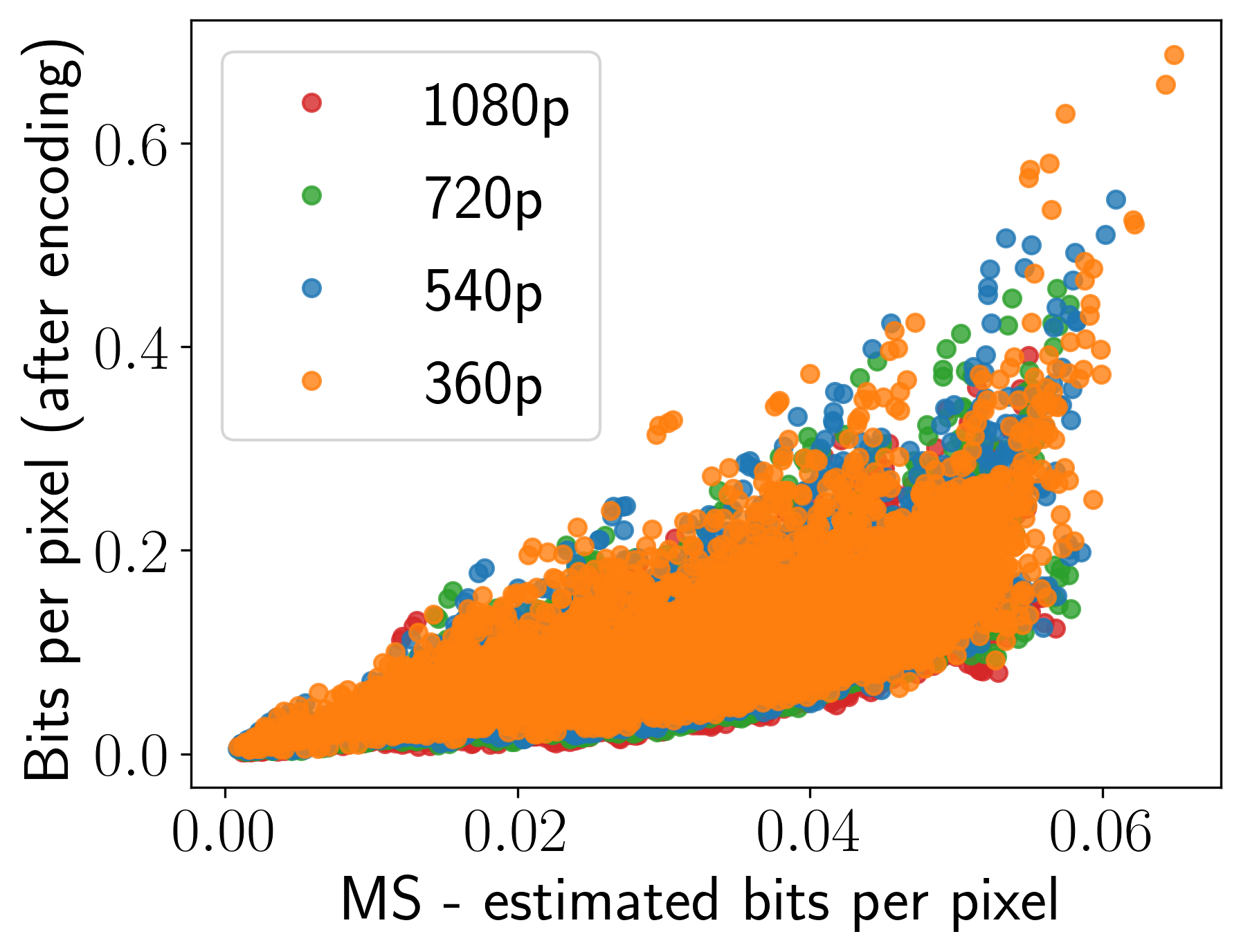}
    \caption{Correlation coefficient: 0.68}
    \label{fig:bitsize-bpp}
    \end{subfigure}
    \begin{subfigure}[b]{0.245\textwidth}
        \includegraphics[width=\textwidth]{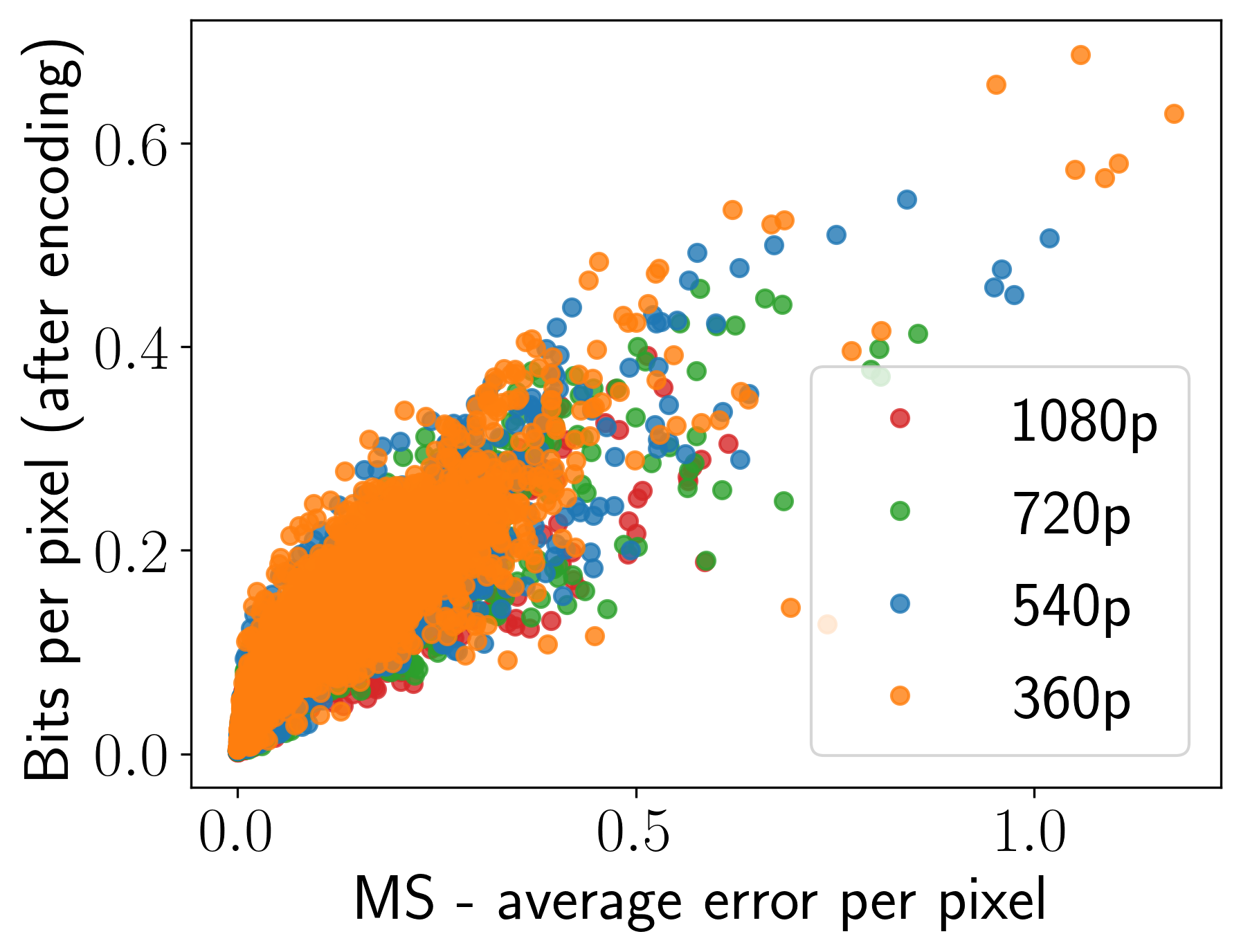}
    \caption{Correlation coefficient: 0.88}
    \label{fig:error-bpp}
    \end{subfigure}
    \begin{subfigure}[b]{0.245\textwidth}
    \includegraphics[width=\textwidth]{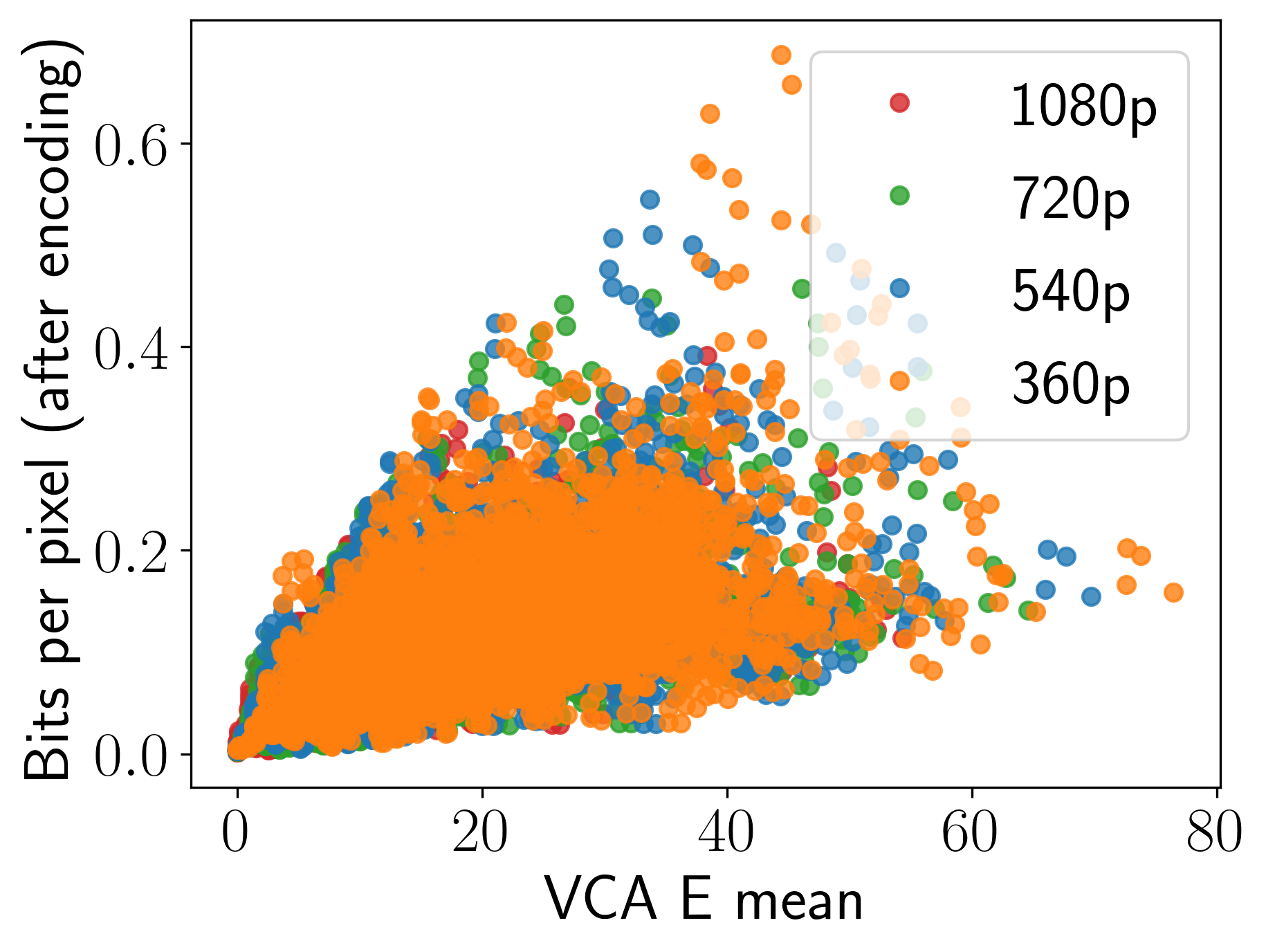}
    \caption{Correlation coefficient: 0.61}
    \label{fig:vca-E-bpp}
    \end{subfigure}
    \begin{subfigure}[b]{0.245\textwidth}
    \includegraphics[width=\textwidth]{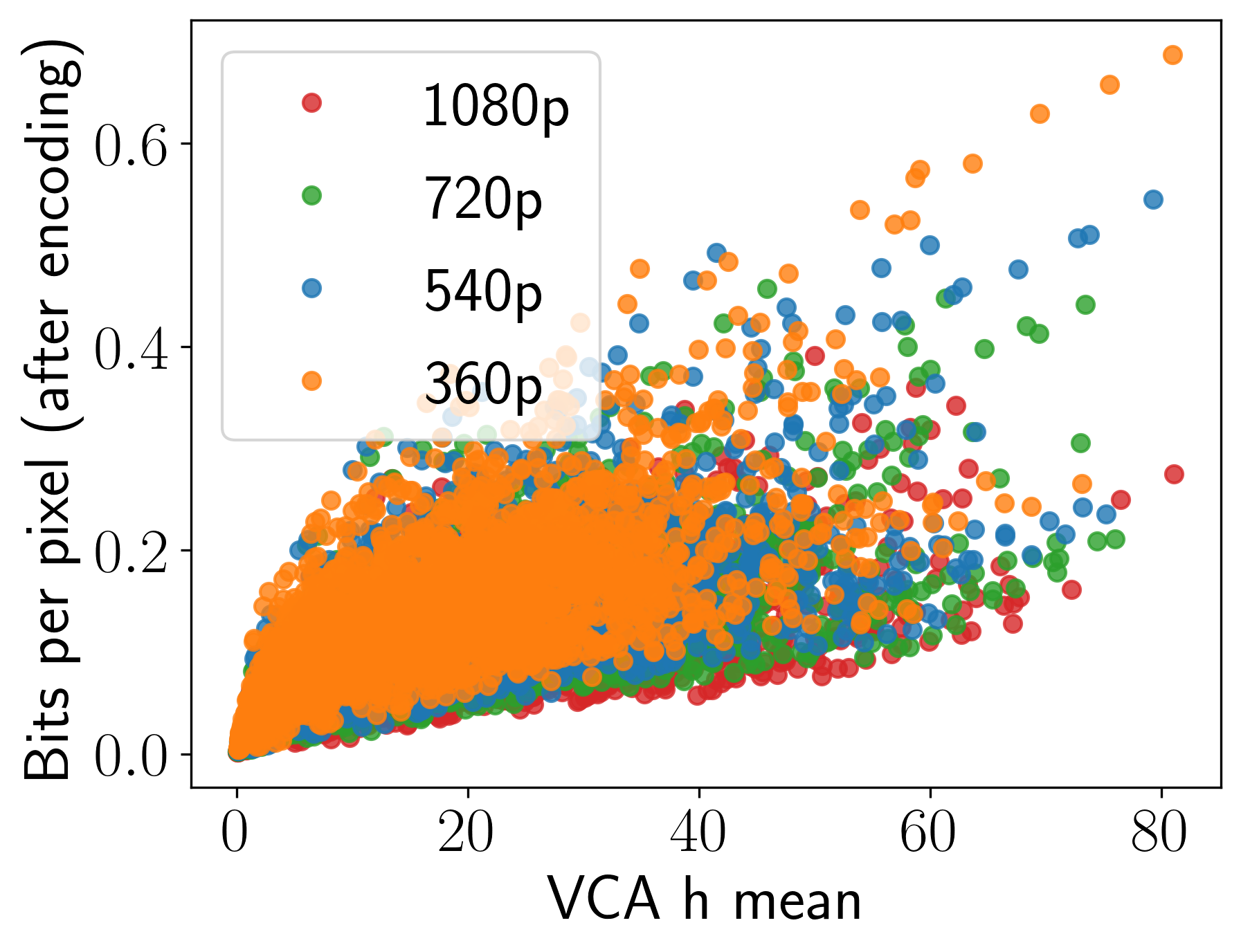}
    \caption{Correlation coefficient: 0.69}
    \label{fig:vca-h-bpp}
    \end{subfigure}
    \caption{Correlation between encoded bits per pixel and estimated bits per pixel (a) and mean squared error between motion vectors (b). Also, the correlation between spatial (c) and temporal (d) complexity obtained from VCA is shown. All show correlation, to some extend, to the encoded bits per pixel. Furthermore, we observe that the normalized values are independent of the number of pixels (i.e. the resolution). Encoded bits per pixel are obtained for preset 5 @CRF 32.}
    \label{fig:correlation_plots}
\end{figure*}

\subsection{Correlation to Bitrate}
We measure the correlation of the descriptors obtained from \eqref{eq:mse-ms} and \eqref{eq:mse-bpp} between the bitrate measured in bits per pixel (bpp).
The results are shown in Fig.~\ref{fig:correlation_plots}. 
We see that especially for $\mathrm{MSE}_\mathrm{MS}$ a high correlation of $0.88$ can be seen, that is specifically higher as the correlation for VCA ($0.61$ for spatial and $0.69$ for temporal complexity).
Correlation is measured in terms of the Pearson Correlation Coefficient (PCC)
~\cite{pearson1896vii}.
\section{Dataset}
We use a subset of the Inter4K dataset~\cite{stergiou2021adapool}, specifically the downsampled videos with resolutions $1080$p, $720$p, $540$p, and $360$p at framerates $24,30,50,$ and $60$. 
The Inter4K video sequences consist of non-pristine online videos, i.e. they have already been encoded. 
Additionally, we want to test on pristine video data and also test the final models on a subset of the AOM Common Test Conditions~\cite{zhao2021aom}. 
From the AOM dataset, we evaluate all horizontal sequences from classes A2, A3, and A4, which consist of $1080$p, $720$p, and $360$p sequences, respectively.
Framerates range from $25$ to $60$.
We use the SVT-AV1 encoder~\cite{kossentini2020svt} to encode the sequences with presets $5$ and $10$ at constant rate factors (CRFS) of $32, 43, 55$, and $63$. We use random access configuration and the default GOP-size of $\sim 5 s$.

\section{Bitrate Prediction Models}
We test a parametric polynomial model and a Random Forest estimator.
For the parametric model, we seek the parameters
$\boldsymbol{\vartheta} = \left( \vartheta_0, \vartheta_1, \vartheta_2, \vartheta_3 \right)^T$ of
\begin{multline}    
    \mathrm{bpp}(\mathrm{bpp}_\mathrm{MS}, \mathrm{MSE}_\mathrm{MS}; \boldsymbol{\vartheta}) \\
    = \vartheta_0 \cdot \left( \mathrm{bpp}_\mathrm{MS} \right)^{\vartheta_1} + \vartheta_2 \cdot \left( \mathrm{MSE}_\mathrm{MS} \right)^{\vartheta_3}
    \label{eq:analytical}
\end{multline}
using least-squares fitting.
The Random Forest is constructed of 50 trees.

We create a model for each preset separately. 
Each model is evaluated using $5$-fold cross-validation with a split of $80 \%$ for fitting and $20 \%$ for testing. 
For preset 10 and preset 5, respectively, we create a Random Forest regressor using descriptors from \eqref{eq:mse-ms} and \eqref{eq:mse-bpp} and VCA spatial and temporal complexity. 
Finally, we obtain the following models:
\begin{enumerate}
    \item Polynomial: analytical model from~\eqref{eq:analytical}\label{enum:poly}
    \item VCA: Random Forest constructed from CRF, spatial, temporal complexity
    \item MS: Random Forest constructed from CRF, Motion Search bitsize, error, and I-P-block ratio
    \item MS-VCA: MS and VCA combined
\end{enumerate}
In contrast to the Random Forest regressor, where the CRF value is an additional parameter of the model, we do not include the CRF value into the equation from model~\ref{enum:poly} but rather construct a polynomial for each CRF separately, leading to four different models here. 
\begin{table*}[]
    \begin{subfloat}[Mean average precision.]{
    \begin{tabular}{|c|c|c|c|c|}
        \hline  
         $\mathbf{\epsilon}$ & VCA~\cite{amirpour2022light} & Polynomial & MS & MS-VCA \\
         \hline
         Preset 5 & $7.2 \%$ & $21.5 \%$ &  $6.2 \%$ & $\mathbf{5.3} \%$ \\
         Preset 10 & $8.6 \%$ & $12.8 \%$ &  $7.0 \%$ & $\mathbf{6.3} \%$ \\
         \hline
    \end{tabular}
    \label{tab:errors_all_mape}
    }\end{subfloat}
\hfill
    \begin{subfloat}[Pearson correlation.]{
    \begin{tabular}{|c|c|c|c|c|}
        \hline  
         \textbf{PCC} & VCA~\cite{amirpour2022light} & Polynomial & MS & MS-VCA \\
         \hline
         Preset 5 & $0.907$ & $0.893$ &  $0.954$ & $\mathbf{0.961}$ \\
         Preset 10 & $0.886$ & $0.950$ &  $0.950$ & $\mathbf{0.961}$ \\
         \hline
    \end{tabular}
    \label{tab:errors_all_pcc}
    }\end{subfloat}
    \caption{Pearson correlation over all models tested on the Inter4K dataset. Best performing models for each preset is in both cases MS-VCA (indicated in bold).}
    \label{tab:errors_all}
\end{table*}
\begin{table*}[]
    \begin{subfloat}[Mean average precision.]{
    \begin{tabular}{|c|c|c|c|c|}
        \hline  
         $\mathbf{\epsilon}$& VCA~\cite{amirpour2022light} & Polynomial & MS & MS-VCA \\
         \hline
         Preset 5 & $20.8 \%$ & $15.7 \%$ &  $\mathbf{11.7} \%$ & $11.9 \%$ \\
         Preset 10 & $26.9 \%$ & $17.4 \%$ &  $12.6 \%$ & $\mathbf{11.2} \%$ \\
         \hline
    \end{tabular}
    }\end{subfloat}
\hfill
    \begin{subfloat}[Pearson correlation.]{
    \begin{tabular}{|c|c|c|c|c|}
        \hline  
         \textbf{PCC}& VCA~\cite{amirpour2022light} & Polynomial & MS & MS-VCA \\
         \hline
         Preset 5 & $0.786$ & $0.921$ &  $0.961$ & $\mathbf{0.964}$ \\
         Preset 10 & $0.859$ & $0.928$ &  $0.981$ & $\mathbf{0.983}$ \\
         \hline
    \end{tabular}
    }\end{subfloat}
    \caption{Mean average precision (a) and Pearson correlation (b) over all models tested on the AOM dataset. Best performing models for each preset is in both cases MS-VCA (indicated in bold).}
    \label{tab:errors_all_aom}
\end{table*}

\section{Evaluation}
First, we compare how using the complexity descriptors obtained from Motion Search compares to using VCA as proposed by~\cite{amirpour2022light}. 
Besides the PCC 
we also measure the mean average precision error~(MAPE)
\begin{equation}
    \epsilon(y, \hat{y}) = \frac{1}{n} \cdot \sum_{i = 1}^{n} \frac{\vert y_i - \hat{y}_i\vert}{y_i}
\end{equation}
of the logarithmically scaled bitrates.
Here, $y$ are the logarithms of the real encoded bitrates and $\hat{y}$ their estimates, $n$ denotes the number of samples of either $y$ or $\hat{y}$ and $i$ the accompanying index.

\begin{figure}
    \centering
    \includegraphics[width=0.4\textwidth]{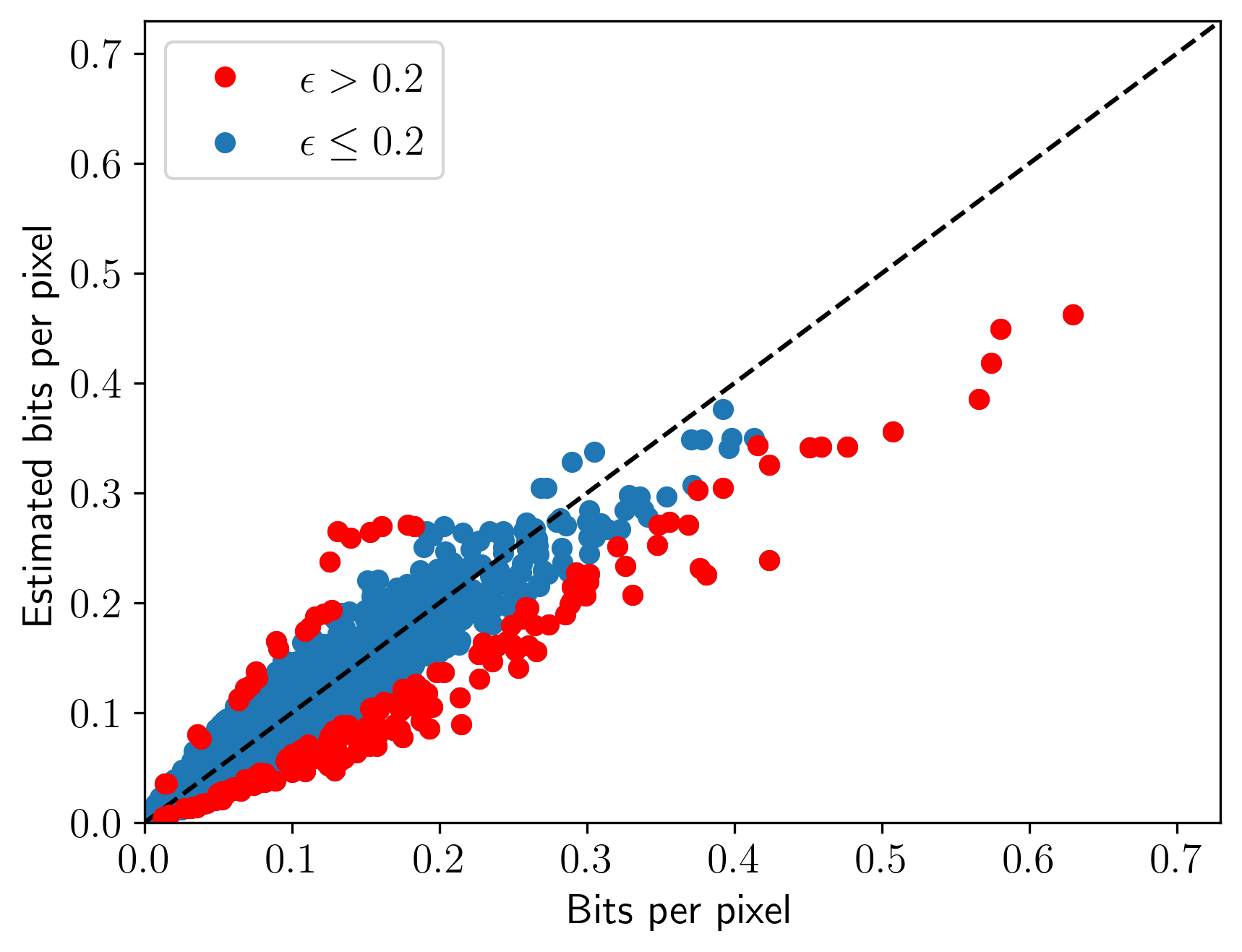}
    \caption{Correlation plot between encoded bits and predicted bits from Random Forest regression using Motion Search and VCA for the Inter4K dataset at preset 5. Relative errors above $20\%$ are indicated by red dots.}
    \label{fig:inter4k-result}
\end{figure}

The performance in terms of MAPE and PCC of the models are concluded in Tab.~\ref{tab:errors_all}.
For the polynomial model, we report the average MAPE and PCC over all CRFs.
Motion Search outperforms VCA when using Random Forest regression and combining both yields the best results. 
The performance of the MS-VCA model is shown in the correlation plot in Fig.~\ref{fig:inter4k-result}.
We also see that the polynomial model performs quite well compared to the Random Forest despite its simplicity.
Even though its accuracy is smaller, it has the advantage to be differentiable, which may be of interest for optimization problems.
\begin{figure}
    \centering
    \includegraphics[width=0.4\textwidth]{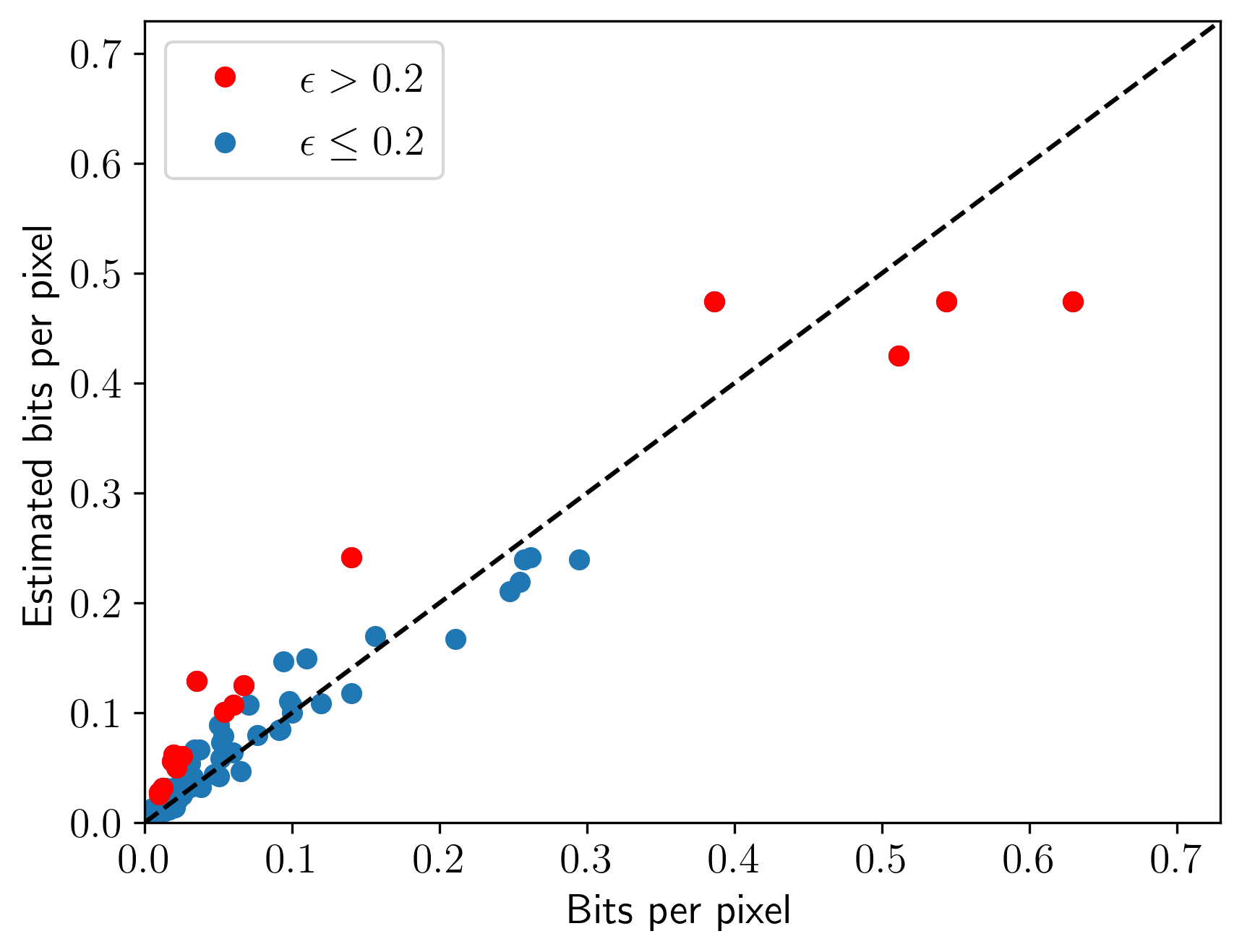}
    \caption{Correlation plot between encoded bits and predicted bits from Random Forest regression using Motion Search and VCA for the AOM dataset at preset 5. Relative errors above $20\%$ are indicated by red dots.}
    \label{fig:aom-result}
\end{figure}
Furthermore, we want to test the performance on pristine data, hence we use the constructed models from the Inter4K dataset to predict the bitrates for encoding the AOM sequences.
The results in terms of PCC are shown in Tab.~\ref{tab:errors_all_aom} and in Fig.~\ref{fig:aom-result}.
Even though we report the same or even higher -- and thus better -- value ranges for PCC, the MAPE increases.
We see in Fig.~\ref{fig:aom-result}, that a much larger number of sequences reaches an prediction error above $20\%$, compared to the total number of sequences.
The reason for this observation may be that for pre-encoded data, as in the Inter4K dataset, spatially varying structure is already smoothed out.
Therefore, the regressor is trained on generally lower spatial variances and falsely classifies the higher variance in seen in pristine data to be more complex to encode and thus overestimated the bitrate here.
\begin{figure}
    \centering
    \includegraphics[width=0.4\textwidth]{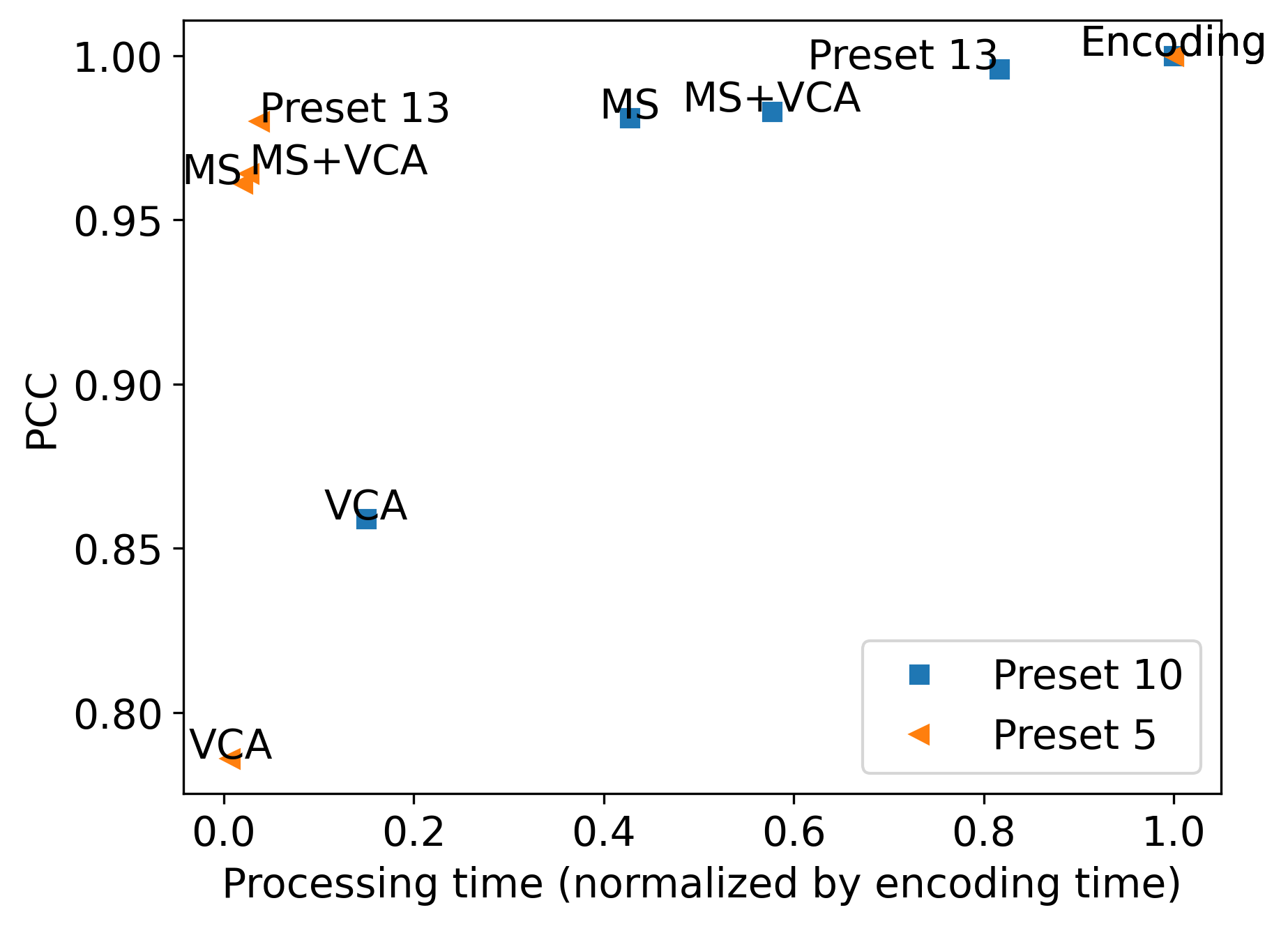}
    \caption{Trade-off between complexity and accuracy measured as PCC for the AOM dataset. Complexity is measured as processing time of the feature, i.e. for VCA, MS, MS-VCA, and encoding with preset 13, divided the time required to encode the sequence (at preset 5 or 10). 
    MS achieves almost the same accuracy as encoding with preset 13, at approximately only half the complexity. VCA has lowest complexity, however, a significantly lower accuracy as well. Overall, MS and MS+VCA require approximately only half the computation time compared to preset 13 while still achieving almost the same accuracy. }
    \label{fig:copmlexity}
\end{figure}

Considering the complexity, we measure the processing time of the encoding process itself, for Motion Search calculation as well as for VCA. Dividing feature time, i.e. CPU time for calculating VCA or performing Motion Search, by encoding time yields VCA being three times faster on average as visualized in Fig.~\ref{fig:copmlexity}. However, both are still significantly faster than encoding itself, i.e. Motion Search takes $47.1 \%$ of the time for encoding with preset 10 and $1.7 \%$ when encoding with preset 5.
Motion Search is also faster than using preset 13, the fastest preset of SVT-AV1.

\section{Conclusion}
Bitrate estimation for AV1-encoding can be achieved using features extracted from Motion Search.
We developed an analytical model that describes the relationship between Motion Search features and encoded bits per pixel with an accuracy almost as high as using machine learning methods.
It outperforms VCA as a feature descriptor in Random Forests and combining both yields the best results for bitrate estimation.
In future work we would like to adapt this scheme to other block-based encoders.
In particular, we seek to improve Motion Search as a complexity estimator, as in reducing its processing time. 

\bibliography{references/bitrate_model}

\end{document}